\documentclass[preprint,11pt]{elsarticle}

\usepackage{graphicx}
\usepackage{amssymb,hyperref}
\usepackage{mathtools}
\usepackage{amsmath,amsthm,amsfonts}
\usepackage[inline]{changes}
\usepackage{graphicx,dsfont}
\usepackage{color}
\usepackage{lineno}
\usepackage{url}
\usepackage[left=1in,right=1in,top=1in,bottom=1in]{geometry}

\usepackage{subcaption}

\def \bu{{\bf u}}
\def \bx{{\bf x}}

\usepackage{lineno}

\usepackage{natbib}

\begin{document}

\begin{frontmatter}



\title{A machine learning approach to using Quality-of-Life patient
  scores in guiding prostate radiation therapy dosing}


\author{Zhijian Yang$^{a,b}$} 
\address[6]{New
  York University, New York NY 10012} 
\address[7]{Current Address: Applied
  Mathematics and Computational Science Program, University of
  Pennsylvania, Philadelphia, PA 19104}

\author{Daniel Olszewski$^{c,d}$} 
\address[1]{Carroll College, Helena
  MT 59625} 
\address[2]{Current Address: Computer, Information Science and
  Engineering Department, University of Florida, Gainesville FL 32611}

\author{Chujun He$^{e}$} 
\address[3]{Smith College, Northampton MA 01063}

\author{Giulia Pintea$^{f,g}$} 
\address[4]{Simmons University, Boston
  MA} \address[5]{Current Address: Department of Psychology, Tufts
  University, Boston MA 02111} 

\author{Jun Lian$^{h}$} 
\address[8]{Department of Radiation Oncology,
  The University of North Carolina, Chapel Hill, NC 27599} 


\author{Tom Chou$^{i}$} 
\address[9]{Depts. of
  Computational Medicine and Mathematics, UCLA, Los Angeles, CA
  90095-1766} 

\author{Ronald Chen$^{j}$} 
\address[10]{Department of Radiation Oncology,
  University of Kansas Medical Center,
  Kansas City, KS 66160} 

\author{Blerta Shtylla$^{k}$} \address[11]{Department of Mathematics,
  Pomona College, Claremont CA 91711 }

%


\begin{abstract}
Thanks to advancements in diagnosis and treatment, prostate cancer
patients have high long-term survival rates. Currently, an important
goal is to preserve quality-of-life during and after treatment. The
relationship between the radiation a patient receives and the
subsequent side effects he experiences is complex and difficult to
model or predict. Here, we use machine learning algorithms and
statistical models to explore the connection between radiation
treatment and post-treatment gastro-urinary function.  Since only a
limited number of patient datasets are currently available, we used
image flipping and curvature-based interpolation methods to generate
more data in order to leverage transfer learning. Using interpolated
and augmented data, we trained a convolutional autoencoder network to
obtain near-optimal starting points for the weights. A convolutional
neural network then analyzed the relationship between patient-reported
quality-of-life and radiation. We also used analysis of variance and
logistic regression to explore organ sensitivity to radiation and
develop dosage thresholds for each organ region.  Our findings show no
connection between the bladder and quality-of-life scores.  However,
we found a connection between radiation applied to posterior and
anterior rectal regions to changes in quality-of-life. Finally, we
estimated radiation therapy dosage thresholds for each organ.  Our
analysis connects machine learning methods with organ sensitivity,
thus providing a framework for informing cancer patient care using
patient reported quality-of-life metrics.
\end{abstract}


\end{frontmatter}

\section{Introduction}
\label{S:1}
Approximately 175 thousand new cases of prostate cancer were reported
in 2019 in the United States \cite{ACS2019}. Depending on age and
cancer stage, first-line treatments include prostatectomy, radiation
treatment, and androgen ablation.  Each of these treatments carries
different side effects. For some patients, a prostatectomy is followed
by radiation treatment to minimize the possibility of recurrence.
Radiation planning for each patient begins with a CT scan, which is
followed by the demarcation of the prostate (radiation target) and the
surrounding organs (bladder and rectum) by a physician. Each plan is
customized to a patient as there is some flexibility in the spatial
dosing of radiation with the primary consideration being delivery of
sufficient dosage of radiation to the target organ without
overexposing and damaging surrounding organs and structures.
Radiation treatment (RT) plans are developed using Dose Volume
Histograms (DVH). DVH discard all organ-specific spatial information
and they are usually based on a single planning CT scan that does not
account for anatomical variations over the course of several weeks of
therapy \cite{Marks2010}. Various metrics have been developed in order
to translate the information from a DVH into a computed probability of
uncomplicated tumor control using normal tissue complication
probability models (NTCP) \cite{Marks2010}. These efforts are
necessary as the relationship between exposure to radiation of the
surrounding organs/structures and the severity and probability of
toxicity (urinary and bowel) is still not fully understood. From a
physiological perspective, it is not clear how radiation dosage
affects tissues, organs, as well as their control and function.

In order to develop a radiation plan that minimizes patient side
effects, one needs to quantify these side effects post-radiation.
Traditionally, physician-assessed scoring systems have been widely
used for measuring patient side effects following cancer treatment
\cite{Basch2010}. However, more recent evidence indicates that
clinicians can downgrade the frequency and severity of patients'
treatment-related symptoms
\cite{Basch2010,Pakhomov2008}. Patient-reported health-related
quality-of-life (QOL) surveys are becoming an important tool in
measuring outcomes after cancer treatment
\cite{Sloan2007,Wagner2007}. For example, a study by K. Diao
\textit{et al.}  \cite{Chen2017} explored urinary and bowel symptom
development during treatment using patient-reported QOL scores (from
1, indicating no symptoms, to 5, indicating high frequency of
symptoms).  An IRB waiver was received at the University of North
Carolina for this retrospective study using anonymized data.  The
results showed that average scores progressively increased from
baseline throughout treatment, but all symptoms resolved to baseline
levels by follow-up.  In the context of RT, NTCP models can be
correlated to patient-reported QOL data, as was done in
\cite{Chen2018}. However their analysis focused only on urinary
symptoms during post-prostatectomy radiotherapy and NTCP models rely
on already reduced DVH information. Given the rich organ-specific
information obtained during treatment planning, it could be desirable
to more directly connect 3D patient CT scans and dosing with QOL
scores.


Machine learning methods have become state of the art in many
applications with impressive results; deep convolutional networks have
many times outperformed traditional methods for diagnosis
\cite{Rajpurkar2017} or visual recognition tools
\cite{Krizhevsky:2017}. In biomedical imaging, deep convolutional
neural network (CNN) algorithms have been applied to a wide array of
problems. Of particular interest in our context are medical image
segmentation machine learning algorithms, such as the U-net
\cite{Ronneberger2015} architecture that focuses on semantic
segmentation of biomedical images. The U-net architecture has
appealing features that we will employ in our approach such as it uses
a large number of max-pooling operations to allow for the
identification of global, non-local features and up-convolution to
return images to their original size. In the work of Nguyen \textit{et
al.} \cite{JianDose2018}, a modified U-net architecture was shown to
accurately predict voxel-level dose distributions for
intensity-modulated radiation therapy for prostate cancer
patients. These prior studies indicate tremendous promise of guidance
of radiation treatment planning with artificial intelligence-based
algorithms. Nevertheless, these algorithms can be of significant use
in treatment planning if they can also incorporate QOL predictions
that can provide immediate guidance for the dosimetrist during
clinical plan optimization. To our knowledge, there are no prior
approaches that integrate QOL in machine learning algorithms in the
context of RT for prostate cancer. 


In the department of Radiation Oncology at the University of North
Carolina, QOL scores were collected using a validated questionnaire
\cite{Talcott2006} administered during weekly treatment visits as part
of the routine clinical work-flow for prostate cancer patients. While
QOL data have been studied {\it post} prostate cancer radiation
treatment, data collected {\it during} the course of treatment can
convey important information about symptom development, which can be a
fertile ground for the use of quantitative modeling to guide optimal
RT dosing.  Accordingly, in this paper we analyze data from a
14-question quality-of-life prostate cancer patient survey that was
collected over the span of five years (2010-2015).  The data we
examined tracked patient urinary and bowel side effects before and
along the course of their treatment (about seven
weeks).  Associated with each patient's QOL, we also
examined associated anatomical CT scans and radiation dosing patterns
for approximately 50 patients.


In this paper, we propose a CNN algorithm to explore the connection
between the spatial distribution of the RT dose and the QOL outcomes
reported from patients in our data set.  A significant problem with
CNN algorithms in our context is the need for a sufficiently large
data set; to resolve this issue, we augmented our patient data sets
using interpolation algorithms that generated synthetic patients by
combining existing patient data.  In addition, we used transfer
learning in order to improve the performance of our CNN algorithm and
implemented steps to avoid overfitting the problem. A key goal for our
study was to generate insight into the most radiation-sensitive tissue
regions.

As a comparative alternative to the CNN approach, we also used
analysis of variance and logistic regression to explore organ
sensitivity to radiation and develop dosage thresholds for each organ
region. We identified regions of the rectum that were highly
correlated with changes in individual patient symptoms. Finally, we
estimated radiation therapy dosage thresholds to determine how high
radiation therapy dosage needed to be in order to trigger collateral
symptoms. Combining results from machine learning and direct analyses
of organ sensitivity provides a powerful framework to inform patient
care in the quality-of-life context.

This paper is organized as follows. In the Methods section, we
formulate convolutional neural network algorithms and statistical
models. In the Results section, we demonstrate that the CNN algorithm
we developed can identify correlations between bowel-related symptoms
and radiation. Furthermore, we support these findings through
statistical analyses that explore organ sensitivity to radiation
dosage. Finally, in the Discussion and Conclusions section, we compare
our results with those of previous studies.

\section{Methods}


\subsection{Quality-of-life data}
Our patient-reported data were extracted from a patient
quality-of-life survey containing answers to 14 questions, seven
questions pertaining to urinary symptoms and seven concerned with
bowel symptoms. The specific survey questions and possible responses
are shown in Appendix A. Patients took the survey before undergoing
radiation therapy, once a week during the seven-week-long treatment,
and after completing therapy.  Answers by patient $j$ to question $i$
at time point $t=\{0,1,2,3,4,5,6,7\}$, $a_{ij}(t)$, were scored on a
discrete scale ranging from 1 to 5 (a score of 1 indicating the least
severity in symptoms, and a score of 5 indicating very high severity
in symptoms). We used this quality-of-life survey and de-identified CT
scans and treatment plans for a total of 52 patients (note that 57
patients were provided, however, five patients had incomplete data and
were discarded in our analysis).

The answers $a_{ij}(t=0)$ to the first survey taken before treatment
were used as the baseline of symptoms before radiation
therapy. Subsequent answers $a_{ij}(t\geq 1)$ provide information on
patient $j$'s symptoms associated with question $i$.  In order to
reduce the dimensionality of the data set collected over several
weeks, we developed a single score for urinary-related symptoms and
bowel-related symptoms as follows.

For each patient, we divided the questions and associated answers into
those concerning urinary or bowel function and then identified the
worst (maximum) score $a^{*}_{ij}\equiv {\rm max}_{t}a_{ij}(t)$ a
patient reported throughout the multi-week treatment for each
question.  A single {\it total difference score} for each patient $j$,
$\Delta_{j}$, was computed as the difference $a_{ij}^{*}-a_{ij}(0)$,
summed over the urinary- or bowel-subset of answers $i$:

\begin{equation}
\Delta_{j}^{\rm u}=\sum_{i=1}^{7} \left[a^{*}_{ij} - a_{ij}(0)\right],
\quad \Delta_{j}^{\rm b}=\sum_{i=8}^{14} \left[a^{*}_{ij} - a_{ij}(0)\right].
\label{eq:QoL_Sum}
\end{equation}
Here, we have ordered questions and corresponding answers associated
with urinary function as $i=\{1,2,3,4,5,6,7\}$ and the
bowel-associated questions as $i=\{8,9,10,11,12,13,14\}$.  The total
difference score $\Delta_{j}^{\rm u}$ and $\Delta_{j}^{\rm b}$
represent the total change in the quality of life associated with
urinary and bowel function, respectively.

\begin{figure}[!hbt]
\centering
\includegraphics[width=3.2in]{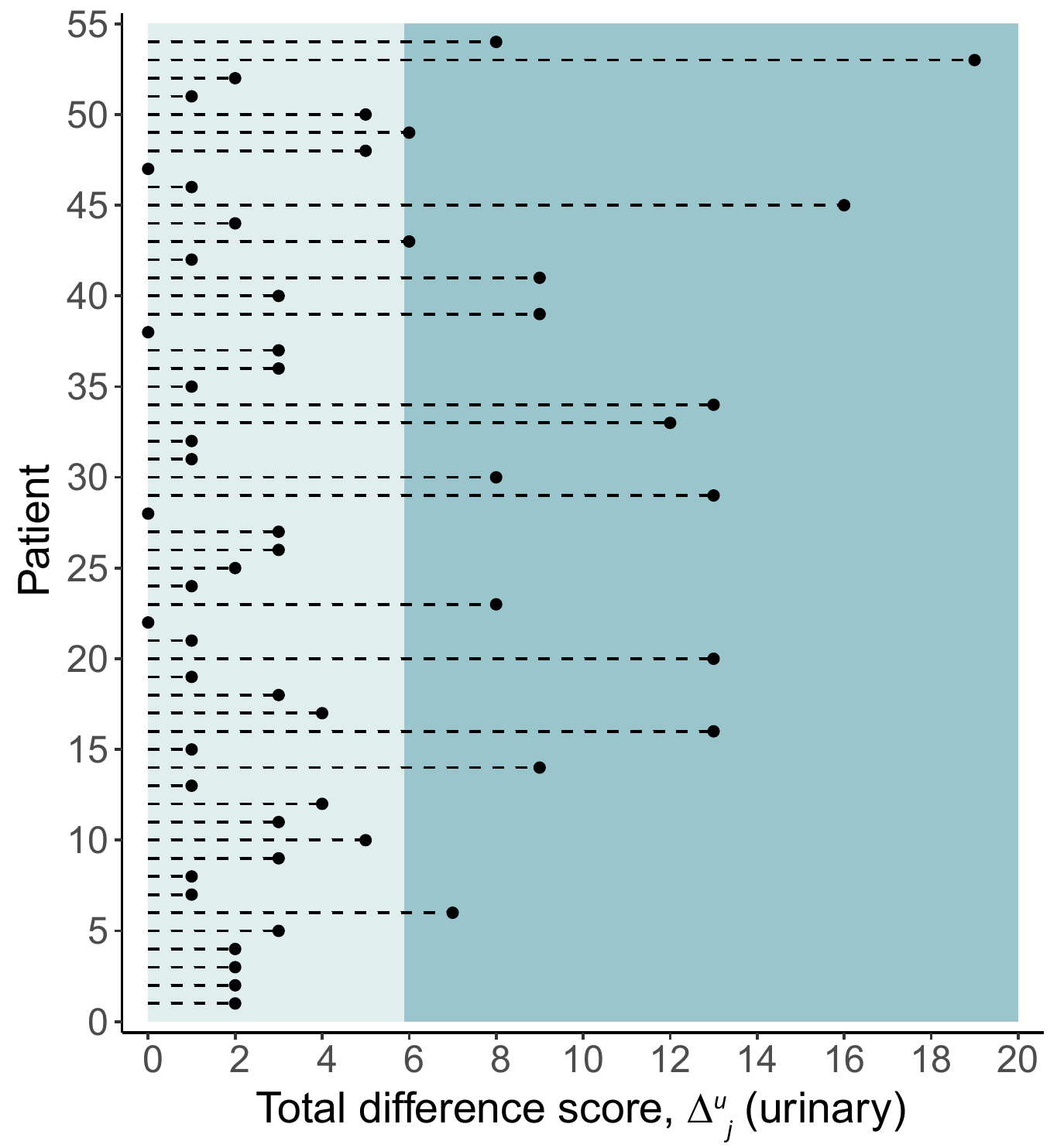}
\caption[Patient Change in Scores]{Total difference scores in urinary
  symptoms for patients 1-54. In blue, we mark patients who were
  classified as having a significant change in symptoms with
  $\Delta_{j}^{\rm u} \geq 6$.} 
\label{fig:binary}
\end{figure}

Next, we used a cut-off (or threshold) value of 6 to convert the patient's
quality-of-life responses into a binary classifier defined {by
  the discrete Heaviside function}

\begin{equation}
y_{j}\equiv H(\Delta_{j}-6) =   \begin{cases} 
      0 & \Delta_{j} < 6, \\
      1 & \Delta_{j} \geq 6.
   \end{cases}\label{eq:binary}
\end{equation}
Score thresholding can be visualized in Figure \ref{fig:binary} that
shows the urinary total difference scores for each patient overlaid
with colored boxes that mark the score cut-off value. The threshold of
6 was initially arbitrarily chosen using the fact that it is more than
half of greatest sum of changes for patients; however, we tested other
thresholds in order to discern issues with the binary classifier as
detailed in the Results section.

\subsection{Image augmentation}
Previous biomedical studies focusing on image processing have used
data sets as small as 85 data points \cite{Kazemifar2018} and as large
as 100,000 data points \cite{Rajpurkar2017}. Since the amount of data
points we have is even smaller than 85, we decided to enrich our
dataset prior to training our algorithms.  We used the
Fischer-Modersitzki (FM) curvature-based image registration technique
\cite{Fischer2004} in order to generate new {\it in silico} patients
as interpolations of existing patient data.


The FM approach was developed in the context of image
registration. For completeness, we outline the core ideas of image
registration algorithms here. Let $d$-dimensional images be
represented by compactly supported mappings $T,R : \Omega \to
\mathbb{R}$ where $\Omega=]0,1[^d$. Specifically, the quantity
    $T(\bx)$ is the intensity or image grey value at the spatial
    position $\bx \in \Omega$. Given a reference image $R(\bx)$ and a
    deformable template image $T(\bx)$, a registration algorithm
    outputs a deformation, or displacement field, $\bu :
    \mathbb{R}^{d} \to \mathbb{R}^d$ such that when applied to the
    template image, $T(\bx-\bu(\bx))$ a resulting modified template
    more closely matches the reference, $R(\bx)$. The problem is then
    how to find a desired deformation $\bu = (u_1, \ldots, u_d)$. This
    becomes an optimization problem as one tries to minimize the
    difference between the deformed template $T_{\bu}\coloneqq
    T(\bx-\bu(\bx))$ and the reference $R(\bx)$. 

Variations in registration methods arise when one employs a particular
optimization technique and one must define a metric for measuring the
goodness of a deformation.  Let $D$ be the distance measure between
the reference $R$ and deformed template $T$, and $S$ a measure of
the smoothness of the deformation $\bu$. The FM approach consists of
finding $\bu$ by minimizing the joint functional $J[\bu]$,

\begin{equation}\label{eq:measureFM}
J[\bu] \coloneqq D[R,T;\bu] + \alpha S[\bu].
\end{equation}
The regularization parameter $\alpha$ is used to control the strength
of the smoothness of the displacement versus the similarity of the
images. The difference-squared 
measure $D$ is given by

\begin{equation}\label{eq:ssdiff}
D[R,T;\bu] \coloneqq {1\over 2} \|R-T_{\bu}\|^2 = {1\over 2} 
\int\limits_\Omega (R(\bx)-T(\bx-\bu(\bx)))^2 {\rm d}\bx, 
\end{equation}
and the curvature-based smoothness

\begin{equation}\label{eq:smoothfunct}
S[\bu] \coloneqq {1\over 2} \sum_{j=1}^d \int\limits_\Omega (\Delta u_j)^2 {\rm d}\bx,
\end{equation}
with Neumann boundary conditions {defined by} 
\begin{equation}
\nabla u_{j}(\bx)=0, \quad \bx \in \partial \Omega, \quad j=1,\ldots,d. 
\end{equation}
We obtain a minimizer $\mathbf{u}$ by first ensuring that the
G\^{a}teaux derivative of the objective function vanishes. The
resulting {\it Euler-Lagrange} equations are

\begin{equation}
f(\bx, \bu(\bx))+\alpha A[\bu](\bx)= 0 , \quad \bx \in \Omega.
\label{euler-lagrange}
\end{equation}
with

\begin{align}\label{eq:distderiv}
f(\bx,\bu(\bx)) &= (R - T_\bu)\cdot \nabla T_\bu(\bx)= 
(R(\bx)-T(\bx-\bu(\bx)))\cdot \nabla T(\bx-\bu(\bx)),\\
A[\bu](\bx) &=\Delta^{2} \bu.
\end{align}
The above semi-linear partial differential equations (PDE) are known
as the {\it Navier-Lame} biharmonic and diffusion equations.


The Euler-Lagrange PDE's can be solved using the following
fixed-point iteration

\begin{align}
\alpha A[\bu^{k+1}](\bx,t)&=-f(\bx,\bu^{k}(\bx,t)), \quad k \geq
 0,\\ \bu^{0}&=0.
\end{align}

Since the computational domain $\Omega$ is of a simple geometry in this case, a
finite difference approximation for the derivatives can be used. This
yields a linear system of equations that are solved in each iteration
step to obtain the deformation $\bu$; more details of the
discretization scheme we implemented to solve for $\bu$ are given in
\cite{Fischer2004}.
 
%

\begin{figure}[!ht]
\begin{subfigure}{.3\textwidth}
\centering
\includegraphics[width=3.5cm,height=3.5cm]{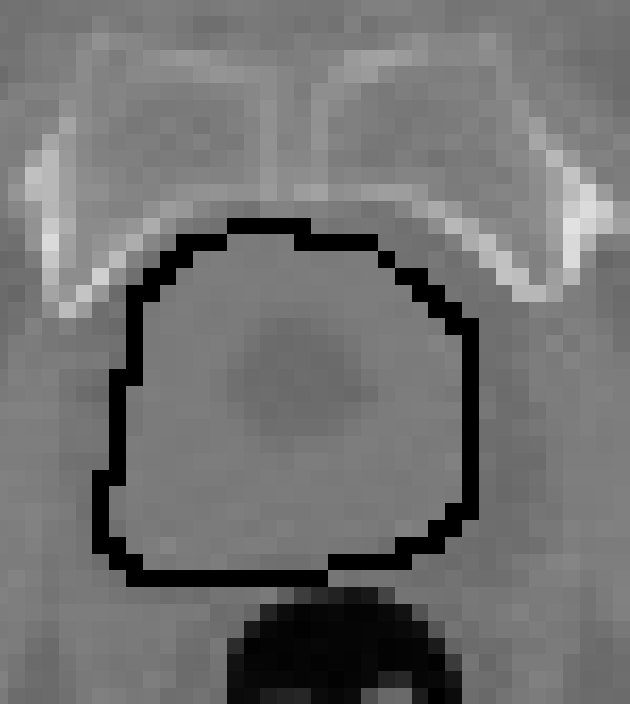}
\caption{Reference}
\end{subfigure}\hfill
\begin{subfigure}{.3\textwidth}
\centering
\includegraphics[width=3.5cm,height=3.5cm]{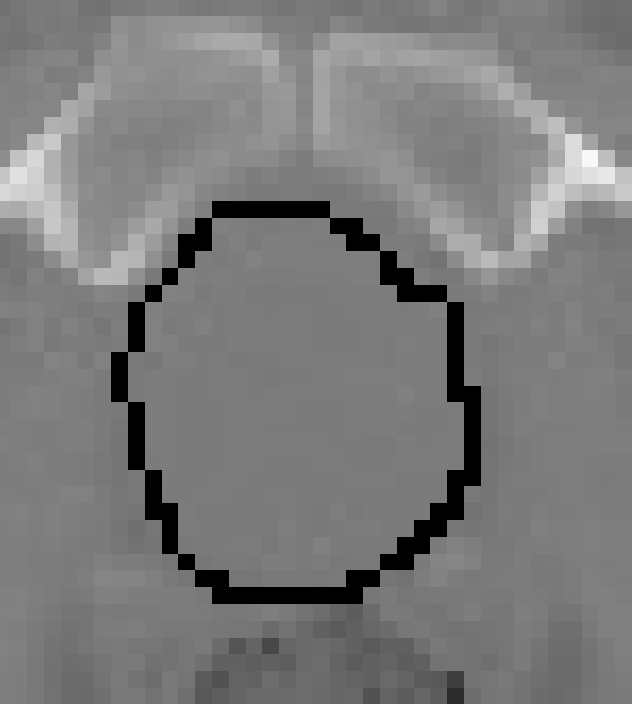}
\caption{Template}
\end{subfigure}\hfill
\begin{subfigure}{.3\textwidth}
\centering
\includegraphics[width=3.5cm,height=3.5cm]{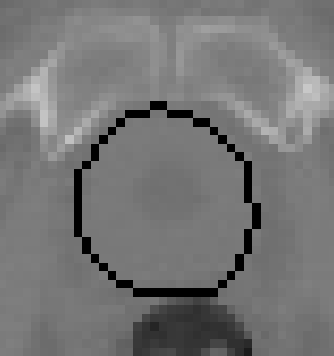}
\caption{Interpolation}
\end{subfigure}
\caption{Interpolation of CT images of the bladder using the FM
  algorithm. The interpolated image (c) is the deformed version of the
  template image (b) against the reference image (a). }
\label{fig:interp}
\end{figure}

Returning to our problem, the goal is to take each CT slice of patient
A and interpolate it with each CT slice of patient B creating a new
stack of CT image slices for a new ``patient'' C. We achieved this
interpolation by applying the FM registration approach and selecting
one stack of images to serve as the reference and another to serve as
the template. We implemented FM in \textsc{Matlab}, using the
discretization approach outlined above and also the implementation
outlined in \cite{MacGillivray2009}. A new interpolated patient C was
obtained by applying the deformation to the template CT image
stacks. We performed this procedure on the 52 patients' cropped CT
image stacks for the bladder and the rectum respectively, thus
producing 1,326 new images for each organ. An example of the
interpolated CT images we obtained using this method is shown in
Figure \ref{fig:interp}.

\subsection{Radiation Plans}

\begin{figure}[!hbt]
\centering
\includegraphics[width=3.5cm,height=3.5cm]{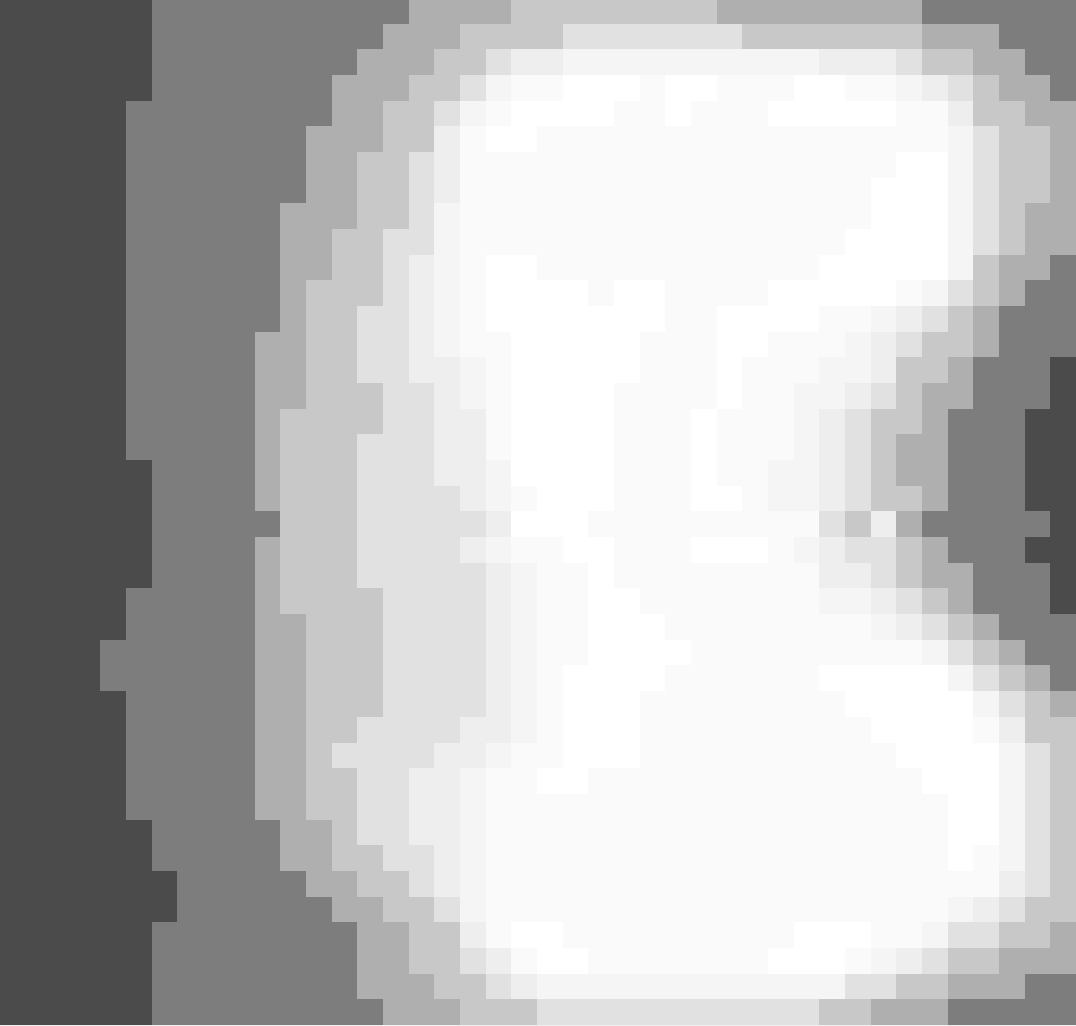}
\caption[Radiation Therapy Plans]{A representative radiation plan. The
  highest dosage corresponds to the greatest pixel intensity (in white). Black
  pixels correspond to no radiation.}
\label{fig:radiation}
\end{figure}

Radiation plans are overlayed on a baseline CT scan and they give the
spatial distribution of the doses that will be given to a patient over
the course of treatment. When the plan is mapped onto a CT image, high
radiation dosage regions are denoted by high pixel intensity (white)
while low dosages are represented by darker pixels.  A
cross-section of a representative radiation plan is shown in Figure
\ref{fig:radiation}. We applied FM interpolation onto the radiation
plan images for each {\it in silico} patient. Specifically, we
interpolate between radiation plans of patient A and patient B to
create a new stack of radiation plans for a new ``patient'' C.


\subsection{Convolutional neural network model}

\begin{figure}[!hbt]
\begin{center}
\includegraphics[width=2.8in]{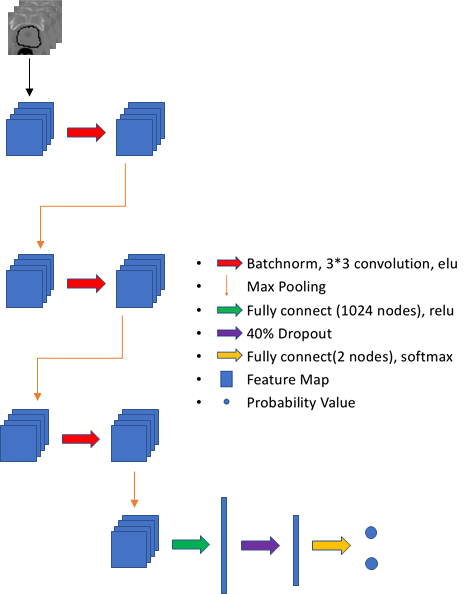}
\caption[Convolutional Classification Architecture]{Architecture used
  for the CNN classification model. There are three layers that have
  convolution, activation, and pooling. The last convolution layer is
  connected by fully connected layers and a drop-out layer which drops
  out 40\% information to avoid over-fitting. Finally the
  model output two nodes which tell the predicted probability that a
  patient will or will not manifest change in (urinary or bowel)
  symptoms throughout treatment}
\label{fig:CNNfigure}
\end{center}
\end{figure}

Next, we constructed a 3-level 3D convolutional neural network (CNN) model
for each organ: one processing data related to urinary symptoms and
one processing data related to rectal symptoms. The architecture of
the model is illustrated in Figure \ref{fig:CNNfigure}.

The convolutional layers use filters of size $3\times 3\times 3$ and
strides of 1. Each of the 3 convolutional layers is followed by max
pooling, which reduced the feature size from $48\times 48$ pixels down
to 1 pixel. We chose to use max pooling for our pooling method so that
the filters captured the strongest (and thus highest) pixel value for
each stride (we use strides of 1 for each filter and a pooling size
of [2,2,2]).  In addition, each convolutional layer is followed by 
an exponential linear unit (ELU) activation function defined as

\begin{equation}\label{eq:ELU}
f(x) = 
  \begin{cases} 
      e^{x} - 1 & x<0, \\
      x & x\geq 0.
   \end{cases}
\end{equation}

We used batch normalization (BN) after the convolution and ELU
operations, which have been shown to update weights equally throughout
the CNN, resulting in faster convergence \cite{Nguyen2018}. In
addition, we used drop-out (40 percent rate) to reduce overfitting
since our dataset was small. Because our
model is a classifier, we use a cross-entropy loss function that the
network minimizes through back-propagation:

\begin{equation}
{\rm Loss}(k) = -\frac{1}{N}\sum_{j=1}^{N}y_{j}
\log p_{k}(y_j) +(1-y_j)\log(1-p_{k}(y_j)),
\label{loss_eq}
\end{equation}
where $k$ is the step number, $y_j = \{0,1\}$ classifier label of the
$j^{\rm th}$ patient, and $p_{k}(y_j)$ represents the predicted
probability for the corresponding label at the $k^{\rm th}$
iteration.

Our CNN had 2 channels as part of its input layer: one to process
information related to the patient CT scans, and one to process
information related to the patient RT plans. The input data consisted
of 52 patient CT scan stacks and RT plans cropped around the organ of
interest (either the bladder or the rectum) with the respective organ
doctor annotated contours for each CT scan. The CT scans and RT plans
were cropped because the information outside of the organs of interest
was not useful for the purposes of this study. Cropping also minimized
the computational power needed to run our algorithm.  The final layer
of the model consisted of two nodes: one providing the predicted
probability that a patient would manifest a change in (urinary or
bowel) symptoms throughout treatment; and one giving the predicted
probability that a patient would not manifest a change in (urinary or
bowel) symptoms throughout treatment. In addition, the CNN produced a
confusion matrix (for either the urinary or bowel symptoms) outlining
how many patients it accurately predicted from the testing set.

\begin{figure}[!hbt]
\centering
\includegraphics[width=5in]{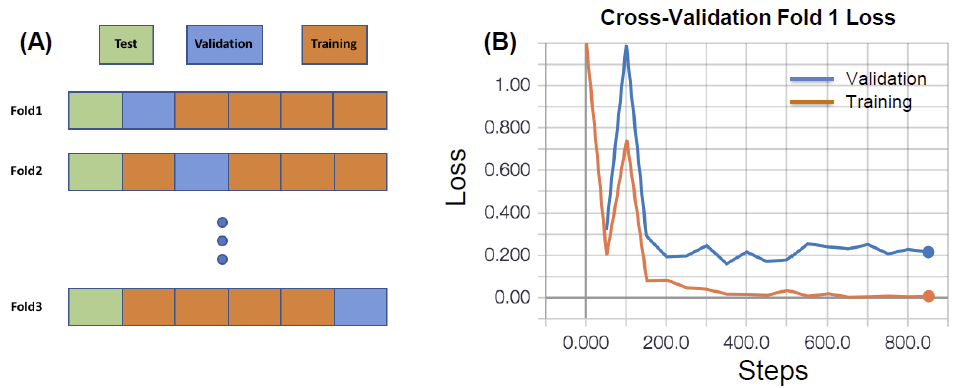}
\caption[Cross-validation Model]{(A) Model cross-validation. We
  initialized 5 different models. For each model there is a validation
  set that moves across the data. First, we trained on the training
  set, then checked each of the models on the different validation
  sets, providing a general idea of which model would work the best.
  (B) Training (orange) and validation (blue) loss for the best model
  chosen through cross-validation.  As shown in orange, the training
  loss becomes very small, but the validation loss stays at about
  $0.2$.}
\label{fig:cross-val}
\end{figure}

To assess the overall performance of our model, the CNN trained on 39
patients with a batch size of 20 and learning rate of 0.001 for
approximately 12 hours, and was then tested against the remaining 13
patients. Patients were shuffled and randomly assigned to the training
and testing sets to avoid bias. The CNN also employed a 5-fold
cross-validation procedure on the training set, similar to the
approach in Jiang \textit{et al.}  \cite{Jiang2015}. Each of the 5
folds split the training set into 31 training patients and 8
validation patients, respectively. Every fold initialized a classifier
(for a total of 5 classifiers), from which we could select the model
that performed best, based on its accuracy and number of
true-positives, and evaluated it on the testing set. We used the
number of true-positives as a criterion for the best-performing model
in order to avoid only predicting a lack of a change in symptoms. The
cross-validation model and corresponding loss functions used can be
visualized in Figure \ref{fig:cross-val}.

\subsection{Autoencoder}
\begin{figure}[!hbt]
\centering
\includegraphics[width=6.2in]{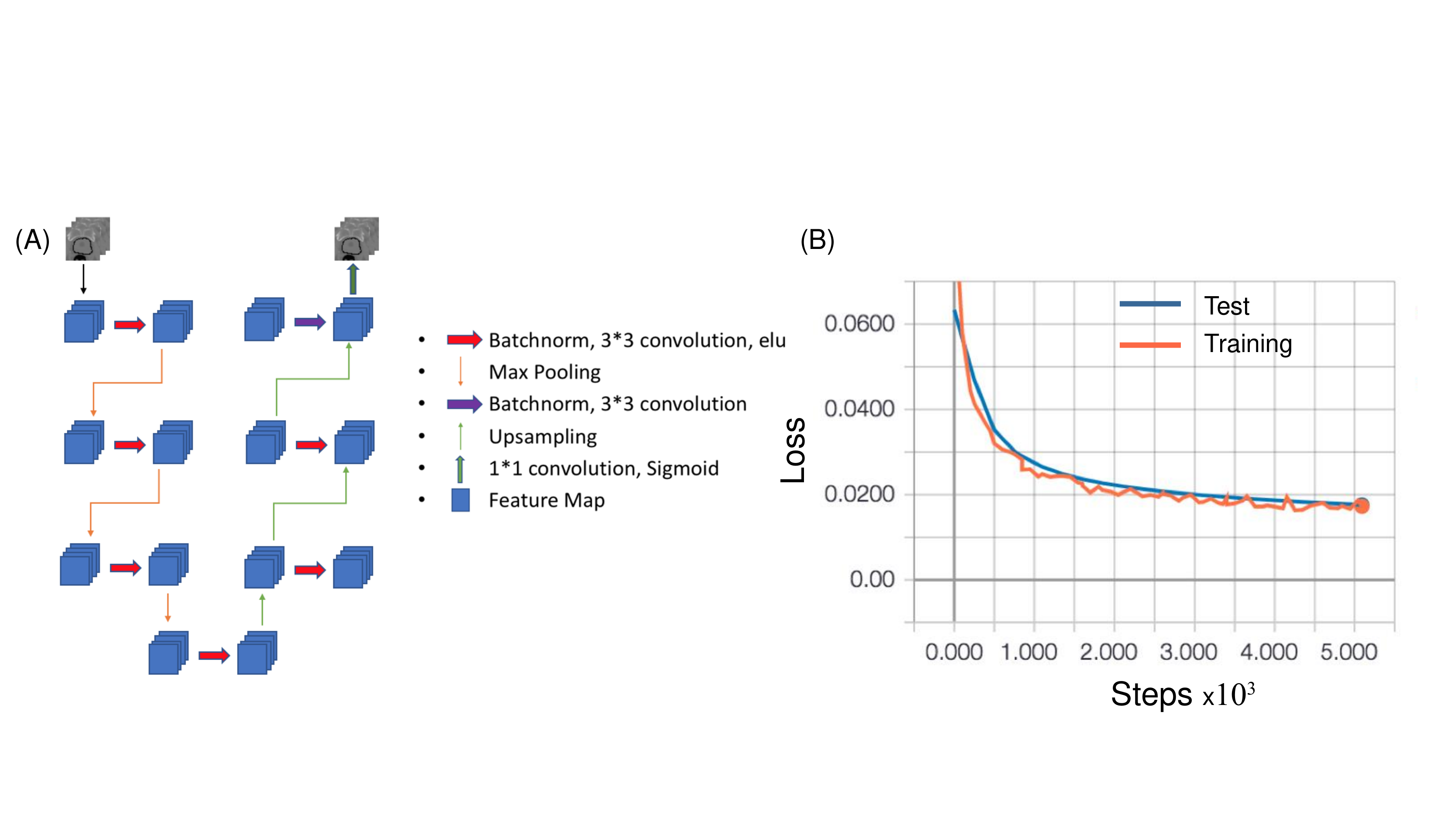}
\caption[Convolutional Autoencoder Architecture]{Convolutional
  Autoencoder. (A) Schematic of the architecture of the convolutional
  autoencoder network. Three layers have convolution, activation, and
  pooling. The network deconvolves with activation and pooling for
  three more layers. The network targets to reconstruct the input
  image with the goal to learn the key features of the CT scans and RT
  plans. (B) Autoencoder loss while training on the bladder data. The
  blue curve shows the loss for the test set and the orange curve
  shows the loss for the training set. Both decrease with the number
  of iterations of the autoencoder.}
\label{fig:autoencoder}
\end{figure}

We employed a convolutional autoencoder network that uses a portion of
the CNN architecture. Specifically, instead of connecting to a fully
connected layer after all the convolution layers as we did with the
CNN, the autoencoder was used to pre-train the network on unlabeled
information by reconstructing the original images. The convolutional
autoencoder network architecture is illustrated in Figure
\ref{fig:autoencoder} and is similar to the U-net architecture used in
related segmentation problems \cite{Kazemifar2018,Balagopal2018}.
After the autoencoder was trained, we truncated the network at the
start of the deconvolution layers and connected it to a fully
connected layer that served as the output for our new CNN model.

Training of the autoencoder allowed us to implement a transfer
learning approach where we first trained the autoencoder network to
reconstruct patient images and assigned the near-optimal weights
obtained from autoencoder training as initial conditions for the CNN
network with the binary classifier. This pre-training was necessary
due to the small size of of the dataset.

\subsection{Statistical analysis for dose thresholding}

In order to further explore the relationship between QOL scores and RT
dosage, we investigated whether or not there were any correlations
between a certain organ region's RT dosage and whether or not the
patient experienced a change in symptoms related to that organ. We
accomplished this through the use of analysis of variance (ANOVA) and
logistic regression.

\begin{figure}[!hbt]
\centering
\includegraphics[width=5.8in,height=1.2in]{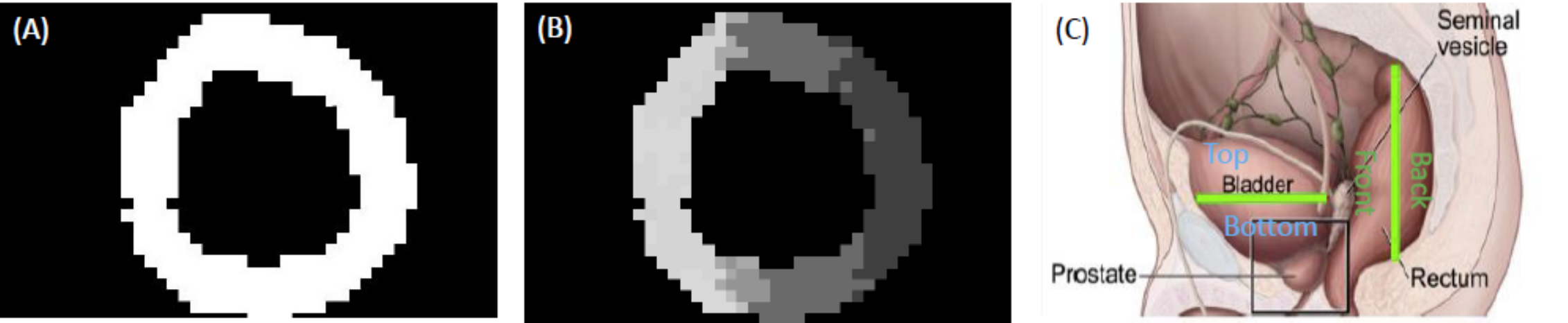}
\caption[Organ Division]{Organ contouring and organ regions. (A) A CT
  slice of the rectum of patient 40, showing the new thick
  contour. (B) Same as (A) but showing the thick contour with the
  corresponding RT dosage. (C) Lateral view diagram showing how the
  bladder (yellow) and rectum (red) were split for spatial RT
  analyses. Image from
  \url{http://libcat.org/anatomy-of-prostate-cancer}.}
\label{fig:OD}
\end{figure}

We first built an algorithm to obtain the RT dosage on the rectum and
bladder contours. We fed the cropped image of the organ of interest
(with overlaid contours from the earlier data preparation process)
into the algorithm and obtained a new image of the organ with thicker
contours. The algorithm was able to do this by locating the center of
the organ and obtaining a radius that traces around the organ's
original (doctor annotated) contour to obtain the new, thicker
contour. This allowed us to obtain more information on the RT dosage
located around the contour of the organ. We were then able to obtain
the RT dosage corresponding to the area occupied by the new, thicker
organ contours. Had we used the original organ contours, we may not
have had the most complete information on the prescribed amount of
RT. The new, thick bladder contours with the corresponding RT dosage
are shown in Figure \ref{fig:OD}.

Following this procedure, we combined all CT images of one patient into
a three dimensional cube and separated it into either top and bottom,
or front and back regions.  We split the bladder into top and bottom
halves and the rectum into front and back regions using the following
rationale. Since the bottom of the bladder is closer to the prostate,
we speculated this region would be exposed to more radiation and thus
be associated with a higher incidence of collateral urinary symptoms.
Therefore, we split the bladder into top and bottom. As for the
rectum, we split it between front and back, since the front is closer
to the prostate and we thus anticipated it would be exposed to a
higher radiation dosage. Organ regions are illustrated in Figure
\ref{fig:OD}(C).

Finally, we define a quantity $d_{lj}$ as the total RT dosage
measured in cGy for a specified organ region, $l$ and patient, $j$.
Total dosages were computed for: ($l=1$) top of the bladder, ($l=2$)
bottom of the bladder, ($l=3$) total bladder, ($l=4$) front of the
rectum, ($l=5$) back of the rectum, and ($l=6$) total rectum.

{\it Organ sensitivity.} We first conducted a paired 2-tailed t-test
to evaluate whether or not the total average dosages were
significantly different across all the regions  outlined above. We
found that all regions had significantly different total RT dosages
($p<0.05$), which further prompted us to investigate whether or not
these differences in RT dosage had an impact on patients' QOL scores
throughout treatment.  To accomplish this, we converted the QOL survey
scores into a binary classifier per question as follows 
\begin{equation}
y^{d}_{ij}\equiv H((a^{*}_{ij}-a_{ij}(0))-2) =   \begin{cases} 
      0 & a^{*}_{ij}-a_{ij}(0) < 2, \\
      1 & a^{*}_{ij}-a_{ij}(0) \geq 2.
   \end{cases}\label{eq:binary_anova}
\end{equation}

  
Using this approach, we obtained a binary value $y^{d}_{ij}$ for each
survey question $i$ and patient $j$ so that we could identify which
symptoms were significantly affected by the corresponding RT
dosage. Then, the binary data were used in one-way ANOVA, which
compared the distribution of RT dosage for a given organ region ($
l=1,\dots,6$) between the two patient groups ($y^{d}_{ij}=1$ vs
$y^{d}_{ij}=0$) for a given symptom question $i$. Recall that using
our convention urinary questions, $i=1-7$ correspond to regions
$l=1-3$ and bowel-associated questions, $i=8-14$ correspond to regions
$l=4-6$.  This analysis allowed us to identify which specific symptoms
were significantly affected by the corresponding RT dosage for a given
organ region.


{\it Dosage Thresholding.}  Since the previous analysis identified
organ regions that were associated with significant changes in
symptoms, we next investigated the ranges of RT dosage that could
trigger such changes. We used logistic regression applied to all the
original binary patient classifiers $y_j$ (from Eq.~(\ref{eq:binary}))
and region total doses, $d_{lj}$ in order to predict whether a patient
would experience changes in symptoms given the corresponding region
dose, $d_{lj}$. The goal of this analysis was to identify the lowest
RT dose for patients with changes in symptoms ($y_j=1$) and the
highest RT dose for patients that did not have a change in symptoms
($y_j=0$).

For each logistic model, a tunable RT dose threshold parameter,
$\theta$ determined to which category patients were assigned based on
the model's prediction, $\hat{y}_{j}$. For each region $l$, we varied
$\theta$ until we obtained a $\theta_{p}$ corresponding to no false
positives and a $\theta_{n}$ corresponding to no false negatives when
comparing $y_{j}$ with the predicted logistic classifier
$\hat{y}_j$. Once we obtained the thresholding parameters for each
region, $l$, we recorded the corresponding RT dosage ranges and used
them to infer how high RT dosage had to be in order to trigger
collateral symptoms.


%



\section{Results and Discussion}

\subsection{CNN results}

We evaluated the performance of the CNN network using a measure of
accuracy defined to be the number of patients with correctly predicted
outcomes over the total number of patients. We estimated accuracy
results for the bladder and the rectum symptoms by running our
algorithm 10 times and averaging the results. We did not find
conclusive results for the bladder, as we obtained an average accuracy
of 38\% with a range of 23\% to 53\%. This indicated that the CNN
model, using the available data, did not find any patterns to classify
the patients for bladder symptoms. For the patients with a change in
bladder symptoms, there was a lot of variability in model predictions
ranging anywhere from 0\% to 50\% with an average of 27\%.

\begin{figure}[!hbt]
    \centering
    \includegraphics[width=3.2in,height=3.in]{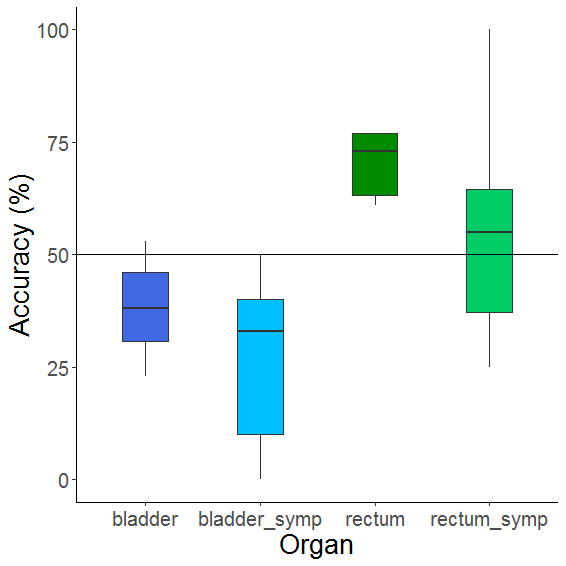}
    \caption{Accuracy for our trained classification model. The
      overall accuracy for the bladder and rectum and the model
      accuracy within patients who experienced symptoms (denoted with
      symp).
%
%
As we can see with the bladder and bladder symptom accuracy, there
were no significant differences, as results sit below the 50\%
line. For the rectum, the overall accuracy for predicting rectum
symptoms exceed the 50\% line.}
    \label{fig:accuracy}
\end{figure}

In contrast, promising results were obtained for classification of
rectal symptoms.  Our model with cross-validation accurately predicted
an average of 74\% changes in rectal symptoms with a range of 62\% to
84.5\%. The results are visualized in Figure \ref{fig:accuracy}.  For
patients with a change in symptoms, the model was on average
accurately predicting the change in symptoms 56\% of the time with a
range from 25\% to 100\%.  In Table \ref{tbl:accuracy}, we give the
confusion matrix for the rectum model with the validation set that
resulted in an 84.6\% accuracy. Of the 10 patients without a change in
symptoms, 9 of them were accurately predicted. Of the 3 patients with
a change in symptoms, 2 were accurately predicted. The result is
promising because the model is not always predicting one of the
classes; it is picking up some patterns from the patients' data so that
it can classify the patients in either category. 

We will show later
that our results are rather insensitive to the threshold delineating
quality-of-life changes.

\begin{table}[!hbt]
\centering
\begin{tabular}{|l|l|l|}
\hline
\multicolumn{3}{|c|}{\textbf{Confusion Matrix (Rectum Symptoms)}}                   \\ \hline
\multicolumn{1}{|c|}{\textbf{Actual}}   & \multicolumn{2}{c|}{\textbf{Predicted}} \\ \hline
                                        & No Change            & Change           \\ \hline
No Change                               & 9                    & 1                \\ \hline
Change                                  & 1                    & 2                \\ \hline
\end{tabular}
\caption[Confusion Matrix for Rectum Model]{Confusion matrix for
  rectum model. Table shows the accuracy for one completely validated
  model. We show the actual classification of the patient, and what
  the model predicted.}
\label{tbl:accuracy}
\end{table}

\subsection{Outcome thresholding}
Other than classifying patients based on their quality-of-life score
with a cut-off value of 6, we also used thresholds of 5 and 7 and used
the reclassified data to train the classification model.
Unsurprisingly, the accuracy rates are similar to those found using a
threshold of 6.

For the data reclassified with a threshold of 5, our model with
cross-validation accurately predicted an average of 69\% changes in
rectal symptoms.

For the data reclassified with a threshold of 7, we found an average
accuracy rate of 69\%.  As there are no considerable differences in
results when we change the threshold, reinforcing that our original
choice of a cut-off-value, the half of the greatest sum of changes, is
a reasonable way to classify patients based on their quality-of-life
scores.

\subsection{Statistical analysis results}
Based on our ANOVA analyses for the bladder data, it does not appear
that there were any significant differences between the two patient
groups in regards to average spatial RT dosing for the bottom and top of the
bladder, as shown in Table \ref{Table_ANOVA_B}. 


\begin{table}[h!]
\centering
\scriptsize
\begin{tabular}{ |l||c|c|}
 \hline \multicolumn{3}{|c|}{ANOVA Dose Difference Results} \\ \hline QoL Question& P-Value (Bottom of Bladder)& P-Value (Top of  Bladder)\\ \hline 
 1(Flow Easiness)&0.5331 & 0.0763 \\ \hline 
 2 (Frequency at Night)&0.2405& 0.8652\\ \hline 
 3 (General Frequency) & 0.58571& 0.687 \\ \hline 
 4 (Pain) &0.6682 &0.1895\\ \hline 
 5 (Urgency) & 0.5272 & 0.094 \\ \hline 
 6 (Control) & 0.9656& 0.1767\\ \hline
 7 (Leak) & 0.7949& 0.8284\\ \hline 
 All Questions & 0.6172&0.6667 \\ \hline
\end{tabular}
\caption[Urinary Symptoms and Average Dosage for Bladder]{P-values
  corresponding to the ANOVA analysis of each urinary symptom and the
  average RT dosage in the top and bottom of the bladder. $*$ indicates
  $p<0.05$ which we interpret as statistically significant.}
\label{Table_ANOVA_B}
\end{table}


\begin{table}[h!]
\centering
\scriptsize
\begin{tabular}{|l||c|c|}
 \hline
 \multicolumn{3}{|c|}{ANOVA Dose Difference Results} \\
 \hline
 Question& P-Value (Front of Rectum)& P-Value (Back of Rectum)\\
 \hline
1 (Diarrhea)&  0.0018* &  0.0002*  \\
\hline
2 (Urgency)&  0.0025*&  0.0006* \\
\hline
3 (Pain)& 0.0023*& 0.0008* \\
\hline
4 (Bleeding)& 0.6563& 0.4142 \\
\hline
5 (Cramping)&  0.0545 &  0.0015* \\
\hline
6 (Pass Mucus)& 0.0387* & 0.0185* \\
\hline
7 (Nothing to Pass)& 0.0541& 0.0504 \\
\hline
All Questions & 0.0123* & 0.0112*\\
 \hline
\end{tabular}
\caption[Bowel Symptoms and Average Dosage for Front of
  Rectum]{P-values corresponding to the ANOVA analysis of each rectal
  symptom and the average RT dosage in the front and back of the rectum.  $*$
  indicates $p<0.05$ which we interpret as statistically significant.}
\label{Table_ANOVA_R}
\end{table}



According to our ANOVA analyses on rectum data, those who experienced
changes in symptoms of diarrhea, urgency, pain, and passing mucus had
significantly higher average RT dosage in the front of the
rectum. Furthermore, those who experienced changes across all rectal
symptoms had a significantly higher average RT dosage in the front of
the rectum. We found the same results for the back of the rectum (with
a difference in cramping which turned out to be significant for the
back but not the front). The corresponding p-values for both organ
regions are listed in Table \ref{Table_ANOVA_R}. The results indicate
that the rectum is more sensitive to radiation therapy, 
matching the results from our CNN model.

\begin{figure}[htbp]
\begin{center}
\includegraphics[width=5.6in]{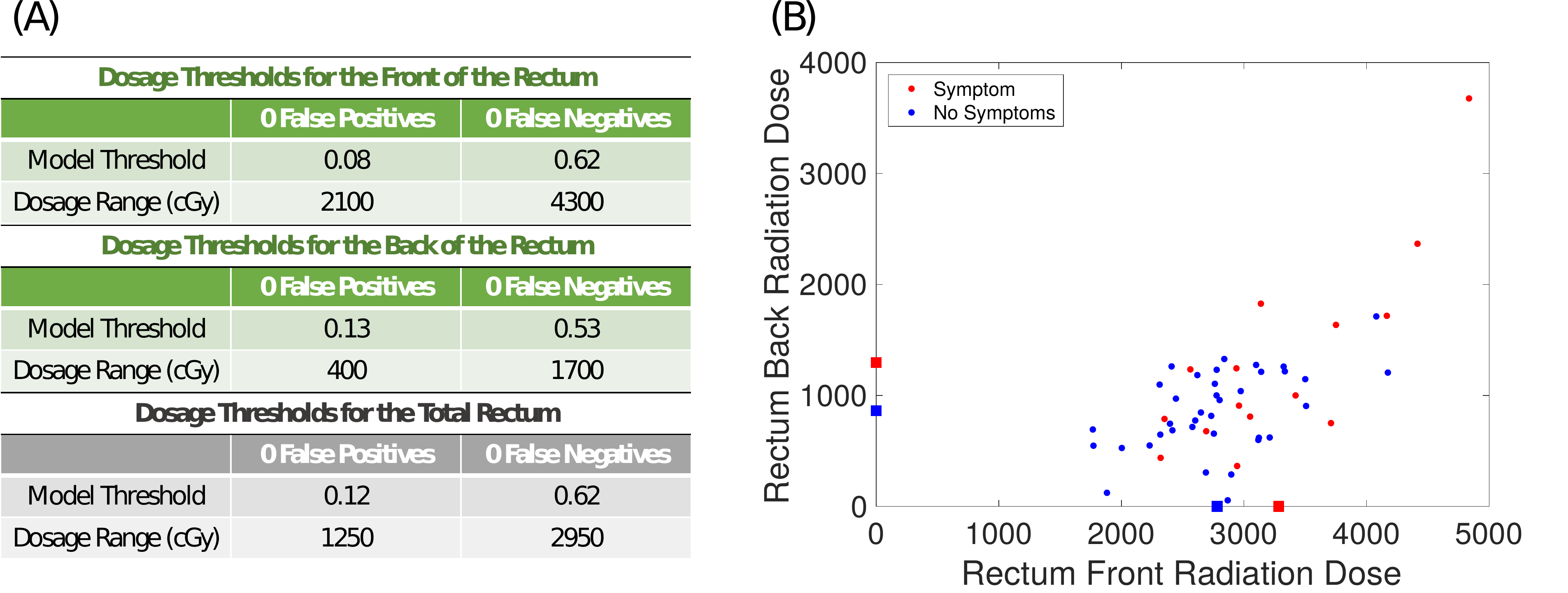}
\caption[Rectum Dosage Thresholds]{Logistic model thresholds and
  corresponding RT dosages for each rectum region. (A) Computed
  thresholds assuming independence. (B) Scatter plot for patients and
  their corresponding front and back RT doses. Patients with symptoms
  and without symptoms are shown in red and blue, respectively. The
  distribution of sampled RT doses are just broad enough to observe
  the that higher doses lead to symptoms. The mean front and back
  radiation doses of patients with and without symptoms are indicated
  by the thick red and blue bars on the $x$- and $y$-axes,
  respectively. The current data are insufficient to resolve anything
  other than a total dosage thresholding effect.}
\label{LR_R}
\end{center}
\end{figure}

According to our logistic threshold analysis, we found that patients
tend to develop bowel-related symptoms throughout the treatment if
they are receiving doses more than 2,500 cGy in the
rectum. Correspondingly, in Figure \ref{LR_R}(A), we also found that
the safe doses to the rectum range from 1,250 to 2,950 cGy, implying
that a dose greater than 2,950 cGy will trigger the development of
collateral symptoms. If we further assume the radiation dosages to the
different parts of the rectum are uncorrelated, we can also
independently find dosage thresholds for the front and back of the
rectum. We observed that the front of the rectum could tolerate a
higher range of absolute dosage (2,100-4,300 cGy) than the back of the
rectum (400-1,700 cGy). These threshold values are listed in Figure
\ref{LR_R}(A) and similar results for the top, bottom, and total
bladder regions are shown in Appendix B.

While this conclusion is consistent with the correlation found between
the rectal symptoms and rectum dosages in the interval 2500-4200 cGy
\cite{Al-Abany2005}, we cannot rule out that it could result from a
possible collinear effect in which patients received high doses to
both front and back of the rectum. Since the symptoms cannot be
associated with excess radiation to the front of back of the rectum,
the region-dependent dosage thresholds are likely to depend on the
dosage experienced in other regions. The dosage-symptom instances are
shown in a scatter plot in Figure \ref{LR_R}(B).  Here, symptoms are
typically associated with larger overall dosages. The data are not
sufficient to resolve independent region-specific thresholds.

For the bladder, the dosage thresholds for the top and bottom of the
bladder showed significant overlap: the top of the bladder could
tolerate a dosage of 0-4,665 cGy while the bottom could tolerate a
dosage of 3,183-5,387 cGy (see Appendix B). The corresponding scatter
plot shows no discernible correlation between symptoms and sampled
dosages.


\section{Summary and Conclusions}

With the lowering of the prostate cancer mortality rate, an emphasis
has been placed on increasing the quality-of-life for patients
undergoing radiation treatment. Utilizing machine learning algorithms
and statistical methods, we provide an in-depth analysis on the
spatial dosage provided to each patient. By analyzing a patient's
anatomical CT image and the radiation therapy dosing, we were able to
connect understanding of how radiation influences secondary
symptoms. We were able to do this by using a convolutional neural
network that analyzed the CT image and associated radiation dosage.
Our second method used ANOVA analysis on summarized spatial
information. Using a brute-force technique, we were able to identify
that splitting the bladder into a top region and bottom and the rectum
into a front and back region was the best approach. Our outcomes from
ANOVA agreed with our convolutional neural network and also provided
dosage thresholds for each region. These results for the dosage
thresholds for the rectum and bladder align with the results we
obtained from our CNN prediction model, but should be interpreted with
care. The thresholds across different regions should not be thought of
as independent parameters because the dosages applied in the patient
samples are correlated and the binary, whole patient symptom
indicators are not attributed to any region. Moreover, the number of
patients and the range of radiation doses they received are not large
enough clearly resolve sharper thresholds. This explains the wide
range for the bladder dosage thresholds and the overlap we observed
between the top and bottom of the bladder.  On the other hand, our CNN
prediction model found that the radiation dosage (and the CT scan
features) do in fact play a large role in explaining the differences
in symptom development across patients.


In conclusion, we developed a deep learning framework and
complementary statistical methods to identify the connection between
spatial dosage and symptoms caused by prostate radiation therapy.  A
strength of machine learning is that it can produce accurate
predictions if presented with sufficiently large data sets; however,
the underlying mechanisms or specific features are difficult to
discern in these approaches. In our application, it has the potential
not only to accurately predict patient side effects, but also to learn
what regions of the organs might be responsible for specific side
effects.  As is significant interest in integrating machine learning
approaches with more traditional modeling approaches, we also found
that classical statistical approaches was also useful in our problem.
We expect that our CNN results will be much more accurate upon
subsequent training on larger patient data sets and can be extended to
predicting specific question scores to further refine treatment
planning.
 
\section{Acknowledgements}
This research was funded by The Jayne Koskinas Ted Giovanis Foundation
for Health and Policy and the Breast Cancer Research Foundation.  TC
acknowledges support from the Army Research Office through grant
W911NF-18-1-0345 and the National Institutes of Health through grant
R01HL146552 (TC). The authors also thank the Institute for Pure and
Applied Mathematics at UCLA for hosting this project under the
Research in Industrial Projects for Students program.




\section{References}

\bibliographystyle{plain} 
\bibliography{UNCC_Bib}

\newpage

\appendix
\section{Quality-of-Life Survey}
\begin{figure}[!hbt]
    \centering
    \includegraphics[width=6.4in]{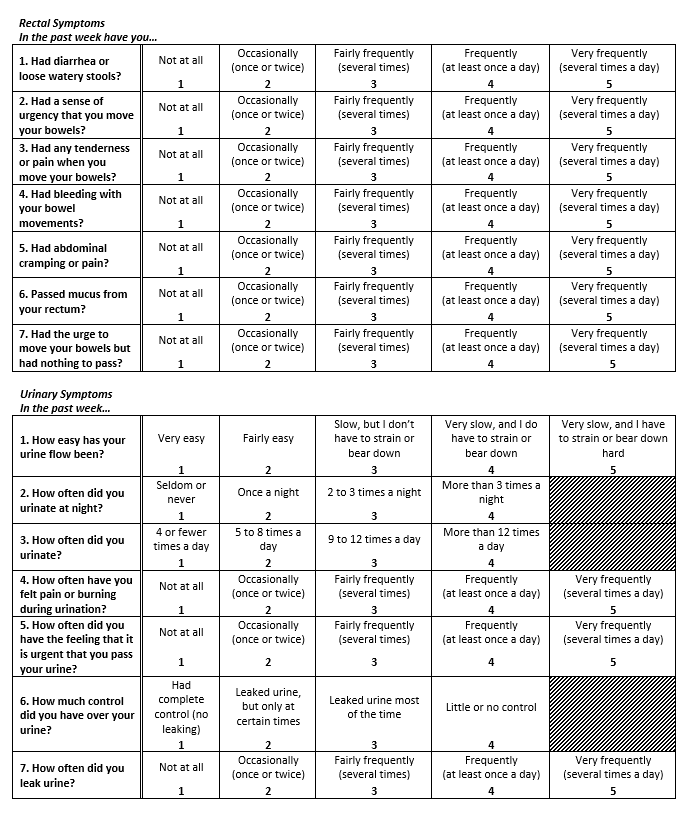}
    \caption{Quality-of-Life surveys given to patients before and 
weekly after RT.}
    \label{fig:survey}
\end{figure}

\begin{figure}[htbp]

\newpage 

\section{Bladder region thresholds}
\begin{center}
\includegraphics[width=5.6in]{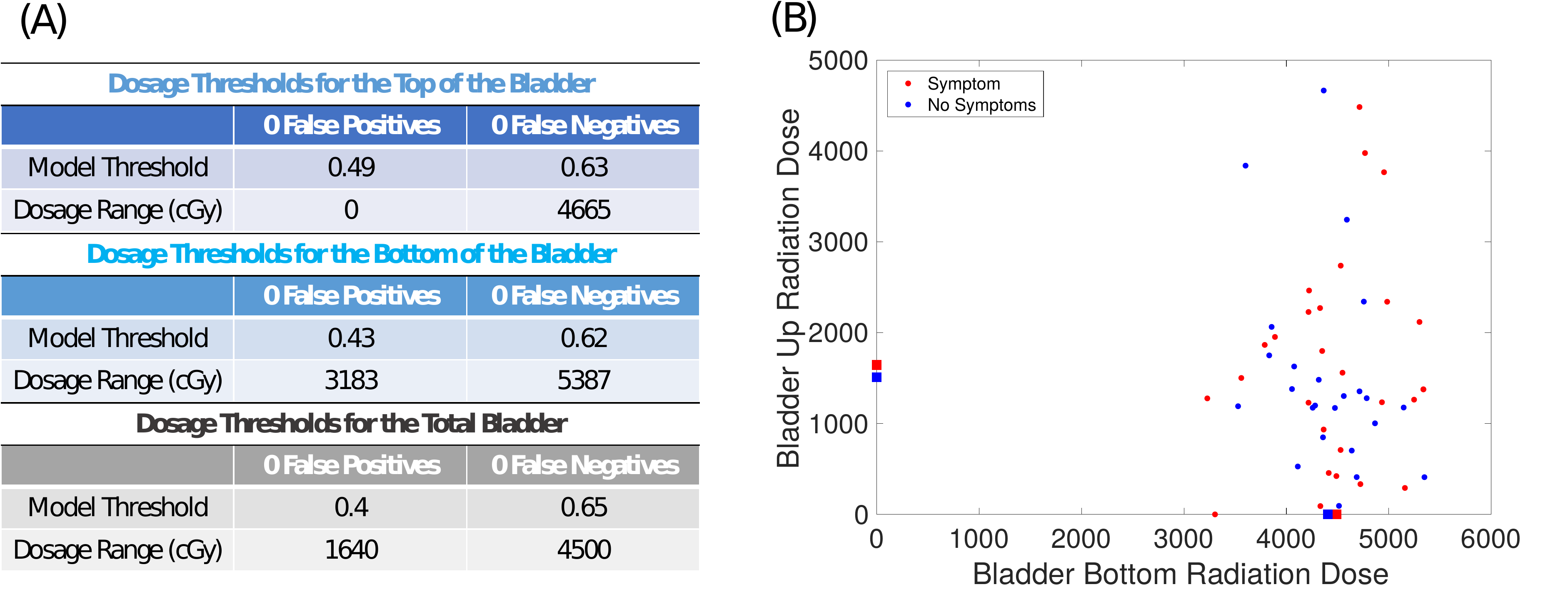}
\caption[Bladder Dosage Thresholds Found with Logistic
  Models]{Logistic model thresholds and corresponding RT dosages for
  each bladder region. (A) Independently computed thresholds.  These
  thresholds were estimated from logistic analysis with significant
  dosage overlap. that contained large overlap. (B) This is made clear
  in the scatter plot of RT doses to the top and bottom regions of
  bladder. For the bladder, there is a much smaller range in bottom
  bladder RT dose and no clear thresholds, i.e., the patients with and
  without symptoms have significant overlap in their RT dosages.}
\label{LR_B}
\end{center}
\end{figure}

\end{document}